\documentclass[nofootinbib,prd,twocolumn,showpacs,showkeys,preprintnumbers]{revtex4}
\usepackage{hyperref,amssymb,amsmath,mathrsfs,bm,graphicx}
\usepackage{epsfig,graphics}
\usepackage{color}
\begin{document}
\title{Relativistic gravitational collapse in comoving coordinates:\\
The post--quasistatic approximation}
\author{L. Herrera \footnote{Also at U.C.V., Caracas.}} 
\email{laherrera@cantv.net.ve}
\affiliation{Departamento de  de F\'{\i}sica Te\'orica e Historia de la Ciencia,
Universidad del Pa\'\i s Vasco, Bilbao, Spain}
\author{W.  Barreto\footnote{On sabbatical leave while beginning this work.}}
\email{wbarreto@ula.ve}
\affiliation{Centro de  F\'{\i}sica Fundamental, Facultad de Ciencias, Universidad de Los Andes, M\'erida, Venezuela}
\date{\today}
\begin{abstract}

A genera iterative method proposed some years ago for the description of relativistic  collapse, is presented here in  comoving coordinates. For doing that we  redefine the basic concepts required for the implementation of the method for comoving coordinates. In particular the definition of  the post--quasistatic approximation in comoving coordinates is given.
We write the field equations, the boundary conditions and a set of ordinary differential equations (the surface equations) which play a fundamental role in the algorithm.  As an illustration of the method, we show how to build up a model inspired in the well known Schwarzschild interior solution. Both, the adiabatic and non adiabatic, cases are considered. 
\end{abstract}
\date{\today}
\maketitle
\section{Introduction}
One of the most outstanding problems in the relativistic astrophysics and gravitation theory today is to provide an accurate description of the gravitational collapse of a supermassive star. The final fate of such process \cite{nbh1}--\cite{nbh8} (naked singularities, black holes, anything else ?), the mechanism behind a type II supernova event  \cite{cw66}--\cite{s7} or the structure and evolution of the compact object  resulting from such a process \cite{e3}--\cite{e2}, stand among the most interesting questions associated to that problem. 

There are essentially two approaches  to describe the gravitational collapse in the context of general relativity. On the one hand, one may   resort  to  numerical methods \cite{lehner}--\cite{font}, which   allow for considering more realistic equations of state. However,  the obtained results,
in general, are restricted and highly model dependents. Also, specific difficulties, associated to numerical solutions of partial differential equations in presence of shocks may complicate further the problem. It would be  desirable in some cases less complicated numerical solvers that community could handle and adapt easily.
On the other hand, one can use analytical solutions to Einstein equations, which  are more suitable for a general discussion, and  may be very useful  in the study of the structure and evolution of self--gravitating systems, since  they may  be relatively simple to analyze but still may contain some of the essential features of a realistic situation. However, often they are found, either for too simplistic equations of state and/or under additional heuristic assumptions whose justification is usually uncertain. Occasionally, analytical approaches \cite{christ1}--\cite{christ5} challenge practical ones, say numeric, allowing discoveries \cite{choptuik1}, \cite{choptuik2} to go further \cite{christ6}, \cite{g99}, \cite{gm07}. Modern numerical relativity without cumulative theoretical insights, would not have developed successfully, of course.

Among many possibilities of exchange between the two aforementioned approaches, some years ago were introduced seminumerical techniques, which  may be regarded as a ``compromise'' between the analytical and numerical approaches. These techniques are based on a general algorithm for modeling self--gravitating spheres out of equilibrium and  were initially developed  for radiation (Bondi) non--comoving coordinates in  \cite{HJR}  (see  \cite{FCP}  for a review and further references). In that version the method has been applied to a variety of different physical scenarios (see for example \cite{hjr2}--\cite{hjr9} and references therein).

Later on the method was extended to Schwarzschild --like coordinates (non--comoving) in \cite{hbds02}, \cite{brm02}. In this format, the algorithm has also lead to a variety of applications  \cite{hjr10}--\cite{hjr13}. Technically, the original method, in radiation coordinates, is first order in the {\it local} radial velocity, whereas in Schwarzschild--like coordinates is second order.

The proposed method (in either version), starting from any interior (analytical) static spherically symmetric (``seed'') solution to Einstein equations, leads to a system of ordinary differential equations for quantities evaluated at the boundary surface of the fluid distribution, whose solution (numerical), allows for modeling, dynamic, self-gravitating spheres, whose static limit (whenever it exists) is the original ``seed'' solution. 

The approach is based on the introduction of a set of conveniently defined ``effective'' variables (effective pressure and energy density) and an heuristic ansatzs on the later, whose rationale and justification become intelligible within the context of the post-quasistatic approximation. In the quasistatic approximation, the effective variables coincide with the corresponding physical variables (pressure and density) and  therefore the method may be regarded as an iterative method
with each consecutive step corresponding to a stronger departure from equilibrium.

It should  be observed, that such  seminumerical techniques require the possibility to approach the non--equilibrium state by means of succesive approximations, implying  that there is life between quasi--equilibrium and  non--equilibrium, at least in some cases.
 
Motivated by the success of previous versions of the method in non--comoving coordinates, and by the fact that comoving coordinates are commonly used in the study of gravitational collapse, we endeavour in this work  to look at some version of the above mentioned algorithm in comoving coordinates, which in turn would require the definition of the post--quasistatic approximation 
in comoving coordinates.

It concerns about the relationship between Eulerian and Lagrangian observers. From the historic point of view, is about Bondi's \cite{b64} and Misner--Sharp's  \cite{ms64} approaches to deal with matter. The former leads to the Wilson codes, the latter to May and White ones \cite{font}, in the context of modern numerical relativity \cite{b09}.

In what follows we develop this plan: First, using comoving coordinates we write down the field equations for the most general fluid. Second, we introduce appropriate definitions of mass and velocity, following Misner and Sharp, to recast the field equations. Third, we detail the junction conditions with the exterior spacetime, which is of Vaidya. Fourth, we consider the static, quasistatic and post--quasistatic regimes. Fifth, we propose a procedure for the modeling and illustrate the algorithm with a simple model based on the Schwarzschild  interior solution.  We consider the adiabatic and the nonadiabatic case. Finally we include some concluding remarks in the last section.

 \section{Comoving frames to describe gravitational collapse}
 \subsection{Comoving coordinates}
 We consider a spherically symmetric distribution  of collapsing
fluid, bounded by a spherical surface $\Sigma$. The fluid is
assumed to be locally anisotropic (principal stresses unequal) and undergoing dissipation in the
form of heat flow (to model dissipation in the diffusion approximation), null radiation (to model dissipation in the free streaming approximation) and shearing viscosity. Physical arguments  to consider such fluid distribution in the study of gravitational collapse may be found in \cite{Herreraanis}--\cite{Ivanov} and references therein.

Using comoving coordinates as in \cite{hsw08}, we write the line element in the form
\begin{equation}
  ds^2=-A^2dt^2+B^2dr^2+R^2(d\theta^2+\sin^2\theta d\phi^2),
\label{metric}
\end{equation}
where $A$, $B$ and $R$ are functions of $t$ and $r$ and are assumed
positive. We number the coordinates $x^0=t$, $x^1=r$, $x^2=\theta$
and $x^3=\phi$.
\subsection{Energy--momentum tensor}
The matter energy--momentum $T_{\alpha\beta}$ inside $\Sigma$
has the form
\begin{eqnarray}
T_{\alpha\beta}&=&(\mu +
P_{\perp})V_{\alpha}V_{\beta}+P_{\perp}g_{\alpha\beta}+(P_r-P_{\perp})\chi_{
\alpha}\chi_{\beta}\nonumber\\
&+&q_{\alpha}V_{\beta}+V_{\alpha}q_{\beta}+
\epsilon l_{\alpha}l_{\beta}-2\eta\sigma_{\alpha\beta}, \label{3}
\end{eqnarray}
where $\mu$ is the energy density, $P_r$ the radial pressure,
$P_{\perp}$ the tangential pressure, $q^{\alpha}$ the heat flux,
$\epsilon$ the energy density of the null fluid describing dissipation in the free streaming approximation, $\eta$ the
shear viscosity coefficient, $V^{\alpha}$ the four velocity of the fluid,
$\chi^{\alpha}$ a unit four vector along the radial direction
and $l^{\alpha}$ a radial null four vector. These quantities
satisfy
\begin{eqnarray}
V^{\alpha}V_{\alpha}&=&-1, \;\; V^{\alpha}q_{\alpha}=0, \;\; \chi^{\alpha}\chi_{\alpha}=1,\nonumber\\ \;\;
\chi^{\alpha}V_{\alpha}&=&0, \;\;\;\;\;\; l^{\alpha}V_{\alpha}=-1, \;\,l^{\alpha}l_{\alpha}=0.\nonumber
\end{eqnarray}

Observe that we have assumed the shear viscosity  tensor $\pi_{\alpha \beta}$ to satisfy the relation \begin{equation}
\pi_{\alpha \beta}=-2\eta \sigma_{\alpha \beta},
\label{sv}
\end{equation}
where   $\sigma_{\alpha \beta}$ is the shear tensor.  However this last equation is valid only within the context of the standard irreversible thermodynamics (see \cite{8}, \cite{FC} for details).

In a full causal picture of dissipative variables we should not assume (\ref{sv}). Instead, we should  use the  transport equation derived from the corresponding theory (e.g. the M\"{u}ller--Israel--Stewart theory \cite{Muller67}--\cite{132}). However for the sake of simplicity, in this work we shall restrict ourselves to the  standard irreversible thermodynamics  theory.

 \subsection{Kinematical variables}
 The four--acceleration $a_{\alpha}$ and the expansion $\Theta$ of the fluid are
given by
\begin{equation}
a_{\alpha}=V_{\alpha ;\beta}V^{\beta}, \;\;
\Theta={V^{\alpha}}_{;\alpha}, \label{4b}
\end{equation}
and its  shear $\sigma_{\alpha\beta}$ by
\begin{equation}
\sigma_{\alpha\beta}=V_{(\alpha
;\beta)}+a_{(\alpha}V_{\beta)}-\frac{1}{3}\Theta h_{\alpha \beta},\label{4a}
\end{equation}
where $h_{\alpha \beta}=g_{\alpha\beta}+V_{\alpha}V
_{\beta}
.$

We do not explicitly add bulk viscosity to the system because it
can be absorbed into the radial and tangential pressures, $P_r$ and
$P_{\perp}$, of the
collapsing fluid \cite{Chan}.

Since we assumed the metric (\ref{metric}) comoving then
\begin{eqnarray}
V^{\alpha}&=&A^{-1}\delta_0^{\alpha}, \;\;
q^{\alpha}=qB^{-1}\delta^{\alpha}_1,  \\
l^{\alpha}&=&A^{-1}\delta^{\alpha}_0+B^{-1}\delta^{\alpha}_1, \;\;
\chi^{\alpha}=B^{-1}\delta^{\alpha}_1, \label{5}
\end{eqnarray}
where $q$ is a function of $t$ and $r$ satisfying $q^\alpha = q \chi^\alpha$.

From  (\ref{4b}) with (\ref{5}) we have for the  four--acceleration and its scalar $a$,
\begin{equation}
a_1=\frac{A^{\prime}}{A}, \;\; a^2=a^{\alpha}a_{\alpha}=\left(\frac{A^{\prime}}{AB}\right)^2, \label{5c}
\end{equation}
where $a^\alpha= a \chi^\alpha$,
and for the expansion
\begin{equation}
\Theta=\frac{1}{A}\left(\frac{\dot{B}}{B}+2\frac{\dot{R}}{R}\right),
\label{5c1}
\end{equation}
where the  prime stands for $r$
differentiation and the dot stands for differentiation with respect to $t$.
With (\ref{5}) we obtain
for the shear (\ref{4a}) its non zero components
\begin{equation}
\sigma_{11}=\frac{2}{3}B^2\sigma, \;\;
\sigma_{22}=\frac{\sigma_{33}}{\sin^2\theta}=-\frac{1}{3}R^2\sigma,
 \label{5a}
\end{equation}
and its scalar
\begin{equation}
\sigma^{\alpha\beta}\sigma_{\alpha\beta}=\frac{2}{3}\sigma^2,
\label{5b}
\end{equation}
where
\begin{equation}
\sigma=\frac{1}{A}\left(\frac{\dot{B}}{B}-\frac{\dot{R}}{R}\right).\label{5b1}
\end{equation}
Then, the shear tensor can be written as
\begin{equation}
\sigma_{\alpha \beta}= \sigma \left(\chi_\alpha \chi_\beta - \frac{1}{3} h_{\alpha \beta}\right).
\label{sh}
\end{equation}

 \subsection{Field equations}
 Thus, the Einstein field equations for the interior spacetime (\ref{metric}) can be written as
 \begin{widetext}
 \begin{eqnarray}
 8\pi\tilde\mu  A^2=\left(2\frac{\dot B}{B} + \frac{\dot R}{R}\right)\frac{\dot R}{R}
 -\left(\frac{A}{B}\right)^2\left[ 2\frac{R^{''}}{R}+\left(\frac{R'}{R}\right)^2
 -2\frac{B'}{B}\frac{R'}{R}-\left(\frac{B}{R}\right)^2\right],\label{12}
 \end{eqnarray}
 \begin{equation}
 4\pi\tilde q AB=\left(\frac{\dot R'}{R}-\frac{\dot B}{B}\frac{R'}{R}-\frac{\dot R}{R}\frac{A'}{A}\right),
 \label{13}
 \end{equation}
 \begin{eqnarray}
 8\pi\tilde P_r B^2=-\left(\frac{B}{A}\right)^2\left[2\frac{\ddot R}{R}-\left(2\frac{\dot A}{A}-\frac{\dot R}{R}\right)
 \frac{\dot R}{R}\right]+\left(2\frac{A'}{A}+\frac{R'}{R}\right)\frac{R'}{R} - \left(\frac{B}{R}\right)^2,
 \label{14}
 \end{eqnarray}
 \begin{eqnarray}
 8\pi \tilde P_\perp R^2=-\left(\frac{R}{A}\right)^2\left[\frac{\ddot B}{B}+\frac{\ddot R}{R} -\frac{\dot A}{A}
 \left(\frac{\dot B}{B}+\frac{\dot R}{R}\right)+\frac{\dot B}{B}\frac{\dot R}{R}\right]
 +\left(\frac{R}{B}\right)^2\left[\frac{A''}{A}+\frac{R''}{R}-\frac{A'}{A}\frac{B'}{B}+\left(\frac{A'}{A}-\frac{B'}{B}\right)
 \frac{R'}{R}\right],
 \label{15}
 \end{eqnarray}
\end{widetext}
where
$$\tilde\mu=\mu+\epsilon,$$
$$\tilde q=q+\epsilon,$$
$$\tilde P_r=P_r-\frac{4}{3}\eta\sigma+\epsilon,$$
$$\tilde P_\perp=P_{\perp}+\frac{2}{3}\eta\sigma.$$
Observe that if functions $A(t,r)$, $B(t,r)$ and $R(t,r)$ are completely determined, the system above becomes an algebraic system of four equations for the six unknown functions $\mu$, $\epsilon$,  $q$, $P_r$, $P_{\perp}$ and $\eta$. In this general  case additional equations are required (e.g. an equation of state and an equation describing energy production) in order to close the system. This is an expression  of the well established fact that under a  variety of  circumstances a line element may satisfy
the Einstein equations for different (physically meaningful) stress-energy
tensors (see \cite{Var}--\cite{varn5} and references therein).
In the locally isotropic case dissipating in either the free streaming or the difussion limit the system is overdetermined. The same happens in the non--dissipative case, even if the fluid is anisotropic.

\section{Mass and Velocity}
Following Misner and Sharp \cite{ms64},
let us now introduce the mass function $m(t,r)$ (see also \cite{cm70}), defined by
\begin{equation}
m=\frac{R^3}{2}{R_{23}}^{23}
=\frac{R}{2}\left[\left(\frac{\dot R}{A}\right)^2-\left(\frac{R^{\prime}}{B}\right)^2+1\right].
 \label{17masa}
\end{equation}
It is useful to define the proper time derivative $D_T$
given by
\begin{equation}
D_T=\frac{1}{A}\frac{\partial}{\partial t}, \label{16}
\end{equation}
and the derivative $D_R$,
\begin{equation}
D_R=\frac{1}{R^{\prime}}\frac{\partial}{\partial r}, \label{23a}
\end{equation}
where $R$ defines the areal radius of a spherical surface inside $\Sigma$ (as
measured from its area).

Using (\ref{16}) we can define the velocity $U$ of the collapsing
fluid  as the variation of the areal radius with respect to proper time, i.e.
\begin{equation}
U=D_TR. \label{19}
\end{equation}
Then (\ref{17masa}) can be rewritten as
\begin{equation}
E \equiv \frac{R^{\prime}}{B}=\left(1+U^2-\frac{2m}{R}\right)^{1/2}.
\label{20x}
\end{equation}

Using (\ref{12})-(\ref{14}) with (\ref{16}) and (\ref{23a}) we obtain from
(\ref{17masa})
\begin{eqnarray}
D_Tm=-4\pi\left[\tilde{P}_rU+\tilde{q}E\right]R^2,
\label{22Dt}
\end{eqnarray}
and
\begin{eqnarray}
D_Rm=4\pi\left(\tilde{\mu}+\tilde{q}\frac{U}{E}\right)R^2.
\label{27Dr}
\end{eqnarray}

Next, the three--acceleration $D_TU$ of an infalling particle inside $\Sigma$ can
be obtained by using  (\ref{14}), (\ref{17masa})  and (\ref{20x}),
producing
\begin{equation}
D_TU=-\frac{m}{R^2}-4\pi\tilde P_r R
+E\frac{A^{\prime}}{AB}, \label{28pce}
\end{equation}
or
\begin{equation}
\frac{A^{\prime}}{A}=\frac{4\pi RB}{E}\left[\frac{D_TU}{4\pi R}+\frac{m}{4\pi R^3}+\tilde P_r\right]. \label{29pce}
\end{equation}

Now, from the Bianchi identities we obtain (see eq. (38) in \cite{hsw08}) in this case
\begin{widetext}
\begin{eqnarray}
&&(\tilde\mu+\tilde P_r)D_T U = \nonumber\\&-&(\tilde\mu + \tilde P_r)
\left[\frac{m}{R^2}+ 4\pi\tilde P_rR\right]
-E^2\left[D_R\tilde P_r+ \frac{2}{R}(\tilde P_r-\tilde P_\perp)\right]-E\left[D_T(\epsilon+q) + 2(\epsilon+q)\left(\frac{2U}{R}+\sigma\right)\right].\label{3m}
\end{eqnarray}
\end{widetext}
The  physical meaning of different terms in (\ref{3m}) has been discussed in detail in \cite{Hs}-\cite{DHN}. Suffice to say in this point that  the first term on the right hand side describes the gravitational force term.

\section{The exterior spacetime and junction conditions}
Outside $\Sigma$ we assume we have the Vaidya
spacetime (i.e.\ we assume all outgoing radiation is massless),
described by
\begin{equation}
ds^2=-\left[1-\frac{2M(v)}{\rho}\right]dv^2-2d\rho dv+\rho^2(d\theta^2
+\sin^2\theta
d\phi^2) \label{1int},
\end{equation}
where $M(v)$  denotes the total mass,
and  $v$ is the retarded time.

The matching of the full nonadiabatic sphere  (including viscosity) to
the Vaidya spacetime, on the surface $r=r_{\Sigma}=$ constant, was discussed in
\cite{chan1} (for the discussion of the shear--free case see \cite{Santos} and \cite{Bonnor}). However observe that we are now including  a null fluid within the star configuration.

Now, from the continuity of the first  differential form it follows (see \cite{chan1} for details), 
\begin{equation}
A dt\stackrel{\Sigma}{=}dv \left(1-\frac{2M(v)}{\rho}\right)\stackrel{\Sigma}{=}d\tau, \label{junction1f}
\end{equation}
\begin{equation}
R\stackrel{\Sigma}{=}\rho(v), \label{junction1f2}
\end{equation}
and 
 \begin{equation}
\left(\frac{dv}{d\tau}\right)^{-2}\stackrel{\Sigma}{=}\left(1-\frac{2m}{\rho}+2\frac{d\rho}{dv}\right), \label{junction1f3}
\end{equation}
where $\tau$ denotes the proper time measured on $\Sigma$.
\begin{widetext}
Whereas the continuity of the second differential form produces
\begin{equation}
m(t,r)\stackrel{\Sigma}{=}M(v), \label{junction1}
\end{equation}
and
\begin{eqnarray}
2\left(\frac{{\dot R}^{\prime}}{R}-\frac{\dot B}{B}\frac{R^{\prime}}{R}-\frac{\dot R}{R}\frac{A^{\prime}}{A}\right)
\stackrel{\Sigma}{=}-\frac{B}{A}\left[2\frac{\ddot R}{R}
-\left(2\frac{\dot A}{A}
-\frac{\dot R}{R}\right)\frac{\dot R}{R}\right]+\frac{A}{B}\left[\left(2\frac{A^{\prime}}{A}
+\frac{R^{\prime}}{R}\right)\frac{R^{\prime}}{R}-\left(\frac{B}{R}\right)^2\right],
\label{j2}
\end{eqnarray}
\end{widetext}
where $\stackrel{\Sigma}{=}$ means that both sides of the equation
are evaluated on $\Sigma$ (observe a misprint in eq.(40) in \cite{chan1} and a slight difference in notation).

Comparing (\ref{j2}) with  (\ref{13}) and (\ref{14}) one obtains
\begin{equation}
q\stackrel{\Sigma}{=}P_r-\frac{4}{3}\eta \sigma.\label{j3}
\end{equation}
Thus   the matching of
(\ref{metric})  and (\ref{1int}) on $\Sigma$ implies (\ref{junction1}) and  (\ref{j3}),
which reduces to equation (41) in \cite{chan1} with the appropriate change in notation. Observe a misprint in equation (27) in \cite{DHN} (the $\sigma$ appearing there is the one defined in \cite{chan1}, which is $-1/3$ of the one used here and in \cite{DHN}).

Also, we have
\begin{equation}
q+\epsilon \stackrel{\Sigma}{=}\frac{L}{4\pi \rho^2}, \label{20lum}
\end{equation}
where $L_\Sigma$ denotes   the total luminosity of the  sphere as measured on its surface and is given by
\begin{equation}
L \stackrel{\Sigma}{=}L_{\infty}\left(1-\frac{2m}{\rho}+2\frac{d\rho}{dv}\right)^{-1}, \label{14a}
\end{equation}
and where
\begin{equation}
L_{\infty} =-\frac{dM}{dv}\stackrel{\Sigma}{=} -\left[\frac{dm}{dt}\frac{dt}{d\tau}(\frac{dv}{d\tau})^{-1}\right]\label{14b}
\end{equation}
is the total luminosity measured by an observer at rest at infinity.

The boundary redshift $z_\Sigma$ is given by
\begin{equation}
\frac{dv}{d\tau}\stackrel{\Sigma}{=}1+z,
\label{15b}
\end{equation}
with
\begin{equation}
\frac{dv}{d\tau}\stackrel{\Sigma}{=}\left(\frac{R^{\prime}}{B}+\frac{\dot R}{A}\right)^{-1}.
\label{16b}
\end{equation}
Therefore the time of formation of the black hole is given by
\begin{equation}
\left(\frac{R^{\prime}}{B}+\frac{\dot R}{A}\right)\stackrel{\Sigma}{=}E+U\stackrel{\Sigma}{=}0.
\label{17b}
\end{equation}
Also observe than from (\ref{junction1f3}), (\ref{14a})  and (\ref{16b}) it follows
\begin{equation}
L\stackrel{\Sigma}{=}\frac{L_\infty}{(E+U)^2},
\label{ju}
\end{equation}
and from (\ref{19}), (\ref{20x}), (\ref{junction1f3}) and (\ref{16b})
\begin{equation}
\frac{d\rho}{dv}\stackrel{\Sigma}{=}U(U+E).
\label{juf}
\end{equation}

\section{Evolution regimes}
We shall next define  three possible regimes of evolution.
\subsection{Static regime}
In this case all time derivatives vanish, implying:
\begin{equation}
\tilde q=U=\Theta=\sigma=0.
\label{1r}
\end{equation}

Since $B=B(r); A=A(r); R=R(r)$, reparametrizing $r$, we  may write the line element in the form:
\begin{equation}
ds^2=-A^2dt^2+B^2dr^2+r^2(d\theta^2+\sin^2\theta d\phi^2).
\label{2r}
\end{equation}

Thus, the ``Euler'' equation (\ref{3m}) becomes the well known TOV equation of hydrostatic equilibrium for an anisotropic fluid
\begin{eqnarray}
P_r^{\prime}
+\frac{2}{r}(P_r-P_{\perp})=-\frac{(\mu+P_r)}{r(r-2m)}
(m
+4\pi P_r r^3).
\label{3r}
\end{eqnarray}
The Einstein equations in this case read:
\begin{eqnarray}
8\pi \mu A^2
=
-\left(\frac{A}{B}\right)^2\left[\left(\frac{1}{r}\right)^2
-2\frac{B^{\prime}}{Br}-\left(\frac{B}{r}\right)^2\right],
\label{4r} 
\end{eqnarray}
\begin{eqnarray}
8\pi P_r B^2 
=\left(2\frac{A^{\prime}}{A}+\frac{1}{r}\right)\frac{1}{r}-\left(\frac{B}{r}\right)^2,
\label{5r} 
\end{eqnarray}
\begin{eqnarray}
8\pi P_{\perp} r^2
=\left(\frac{r}{B}\right)^2\left[\frac{A^{\prime\prime}}{A}
-\frac{A^{\prime}}{A}\frac{B^{\prime}}{B}
+\left(\frac{A^{\prime}}{A}-\frac{B^{\prime}}{B}\right)\frac{1}{r}\right].\label{6r}
\end{eqnarray}

Also, for the mass function we have
\begin{equation}
m=\frac{r}{2}\left(1-\frac{1}{B^2}\right),
\label{7r}
\end{equation}
or
\begin{equation}
B^2=\left(1-\frac{2m}{r}\right)^{-1},
\label{8r}
\end{equation}
or
\begin{equation}
m=4\pi\int^{r}_{0} \mu r^2dr, \label{9r}
\end{equation}
and for the metric function $A$, we have from (\ref{29pce})
\begin{equation}
\ln\left(\frac{A}{A_\Sigma}\right)=\int_{r_\Sigma}^{r}\frac{(m+4\pi r^3 P_r)}{r(r-2m)} dr.\label{10r}
\end{equation}
Therefore, once the radial dependence of $\mu$ and $P_r$ are known, the metric functions are determined  from  (\ref{8r}--\ref{10r}).
\subsection{Quasistatic regime (QSR)}
As  is well known, in this regime the system is assumed to evolve, but sufficiently slow, so that it can be considered to be in equilibrium at each moment (Eq. (\ref{3r}) is satisfied). This means that the sphere changes slowly, on a time scale that is very long compared to the typical
time in which the sphere reacts to a slight perturbation of hydrostatic
equilibrium, this typical time scale is called hydrostatic time scale \cite{astr1}--\cite{astr3} (sometimes this time scale is also referred to as dynamical time scale, e.g. \cite{astr3}). Thus, in this regime  the system is always very close to  hydrostatic
equilibrium and its evolution may be regarded as a sequence of static
models linked by (\ref{13}). 

This assumption is very sensible because
the hydrostatic
time scale is very small for many phases of the life of the star \cite{astr2}.
It is of the order of $27$ minutes for the Sun, $4.5$ seconds for a white dwarf
and $10^{-4}$ seconds for a neutron star of one solar mass and $10$ Km radius.
It is well known that any of the stellar configurations mentioned above, generally (but not always), 
change on a time scale that is very long compared to their respective
hydrostatic time scales. Let us now translate this assumption in conditions
to $U$ and metric and kinematical functions.
This implies that:
\begin{itemize}
\item The areal velocity $U$ as well as other kinematical variables are small, which  in turn implies that  dissipative variables and all first order time derivatives of  metric functions are also small.
\item  From the above and the fact that the system always satisfies the equation of hydrostatic equilibrium, it follows from (\ref{3m}) that second time derivatives of  metric functions can be neglected.
\end{itemize}
Thus in quasi--equilibrium we have to assume:
\begin{equation}
O(U^2) = {\dot A}^2 = {\dot B}^2 =
\dot A \dot B = \ddot R = \ddot B \approx0
\label{11r}
\end{equation}
and the radial dependence of the metric functions as well as that of physical variables is the same as in the satic case. The only difference with the latter case is the fact that variables depend upon time according to equation (\ref{13}).

\subsection{Post--quasistatic regime (PQSR)}
In the two regimes considered above the system is always in (or very close to) hydrostatic equilibrium. Let us now move one step forward into non--equilibrium and let us assume then that (\ref{3r}) is not satisfied. 

Then the question arises: What is the closest situation to quasi--equilibrium, not satisfying eq. (\ref{3r})?
For obvious reasons we  shall call this regime, post--quasistatic regime. Three remarks are in order at this point:
\begin{enumerate}
\item First of all it should be stressed that the main motivation to consider the PQSR is to have the possibility  to consider those aspects of the object directly related to the non--equilibrium situation, which for obvious reasons cannot be described within the QSR.
\item It should be clear that we are also assuming the fact that we can approach the non--equilibrium by means of successive approximations. It goes without saying that not  any self--gravitating fluid will satisfy this requirement.
\item It also should be clear that unlike the two precedent regimes, there is not a unique definition for PQSR. In what follows we shall propose a definition in analogy to  the one given in \cite{hbds02} for non--comoving coordinates.
\item Once the system is out of equilibrium, its eventual return to the static or quasistatic regime  is not assured and will depend on the  initial data and the very nature of the system. 
\end{enumerate}

Now, since in both, the static and quasistatic regimes, the radial dependence of metric variables is the same, we shall keep that radial dependence as much as possible, but of course the time dependence  of  those variables is such that now (\ref{11r}) is not satisfied.

Then from the above we write
\begin{equation}
R=r\kappa(t), \label{pce1}
\end{equation}
where $\kappa$ is an arbitrary function of $t$, to be determined later.

Taking into account (22) and (54), we rewrite the metric as follows
\begin{equation}
ds^2=-A^2dt^2+ \kappa^2[E^{-2} dr^2+r^2(d\theta^2+{\sin}\theta^2 d\phi^2)].
\end{equation}
Next, defining the effective mass as
\begin{equation}
m_{eff}\equiv m-\frac{1}{2}R U^2,
\end{equation}
we obtain
\begin{equation}
E^2=1-\frac{2 m_{eff}}{R}.
\end{equation}
Then, equations (\ref{27Dr}) and (\ref{29pce}) can be written as
\begin{eqnarray}
\frac{1}{\kappa}m_{eff}^{\prime}&=&4\pi R^2 \mu_{eff},\label{drmeff}\\ \nonumber \\
\frac{1}{\kappa}(\ln{A})^{\prime} &=&\frac{4\pi R^2 P_{eff} + m_{eff}/R}{R-2m_{eff}},\label{drA}
\end{eqnarray}
with
\begin{eqnarray}
\mu_{eff}&=&\tilde \mu + \frac{\tilde q U}{E} -\frac{U D_R U}{4\pi R} - \frac{U^2}{8\pi R^2},\label{dense}
\\ \nonumber \\
P_{eff}&=&\tilde P_r + \frac{D_T U}{4\pi R} + \frac{U^2}{8\pi R^2},\label{prese}
\end{eqnarray}
where we have followed  the terminology used in \cite{HJR}, \cite{FCP}, \cite{hbds02} and  call  
$\mu_{eff}$ and $P_{eff}$ the ``effective density'' and the ``effective pressure'', respectively. The meaning of these variables will become clear in the discussion below.

It is worth observing that  in comoving coordinates the PQSR
is ``exactly'' second order in $U$ (or $\tilde q$), which is not the case in non--comoving coordinates (radiation or otherwise).
 
Next, from (\ref{drmeff})--(\ref{prese}), with (\ref{pce1}) we may write
\begin{equation}
\frac{1}{\kappa^3} m_{eff}=\int^r_0 4\pi r^2 \mu_{eff}dr,\label{meff}
\end{equation}
\begin{equation}
\frac{1}{\kappa}\ln\left(\frac{A}{A_\Sigma}\right)=\int_{r_\Sigma}^r \left[\frac{4\pi R^3 P_{eff} + m_{eff})}{R(R-2m_{eff})}\right]
dr.\label{AE}
\end{equation}

From the above, it follows at once that if $R=\kappa(t) r$ and $\mu_{eff}$ shares the same radial dependence as $\mu$ in the static case, then obviously the radial dependence of $m_{eff}$ will be the same as in the static case. The inverse is true of course, if  the radial dependence of $m_{eff}$ is the same as in the static case, then $\mu_{eff}$ shares the same radial dependence as $\mu$ static.

On the other hand, if besides the assumption above, we assume that $P_{eff}$ shares the same radial dependence as $P_{r}$ static, then it follows from (\ref{AE}) that $A$ shares the same radial dependence as in the static case.   

All these considerations provide the rationale for our algorithm. Indeed, starting with a   ``seed'' static solution with a given $\mu(r)$ and $P_{r}(r)$, it follows from (\ref{meff}) and (\ref{AE}),  that the PQSR can be implemented by assuming that the radial dependence of  the effective variables is the same as that of $\mu(r)$ and $P_{r}(r)$ (thogether with $R=r\kappa(t)$).

It is worth stressing an important difference between  the operational definition of QSR and PQSR in comoving and noncomoving frames. In the latter case there is one physical variable more (the velocity) and one metric function less ($R$) than in the former.

We shall next  outline the approach that we propose.

As mentioned before, such an approach was already proposed  and developed in a noncomoving frame of reference (see \cite{HJR}, \cite{FCP} and \cite{hbds02} and references therein). Here we  want to  provide a formulation  for comoving frame.

The proposed method, starting from any interior (analytical) static spherically
symmetric (``seed'') solution to Einstein equations, leads to a system of ordinary differential equations for quantities evaluated at the boundary surface of the fluid distribution,
whose solution (numerical), allows for modeling, dynamic, self-gravitating spheres, whose static limit is the original ``seed'' solution.

The approach is based on the  post-quasistatic assumption, which as mentioned before is equivalent to the assumption that   ``effective'' variables (effective pressure and energy density)  share the same radial dependence as the radial pressure and energy density  of the static ``seed'' solution, respectively.  An ansatzs justified by the fact that  in the quasistatic approximation, the effective variables coincide with the corresponding physical variables (pressure and density).  Therefore the method may be regarded as an iterative method
with each consecutive step corresponding to a stronger departure from equilibrium.
\begin{figure}[htbp!]
\begin{center}
\scalebox{0.4}{\includegraphics[angle=0]{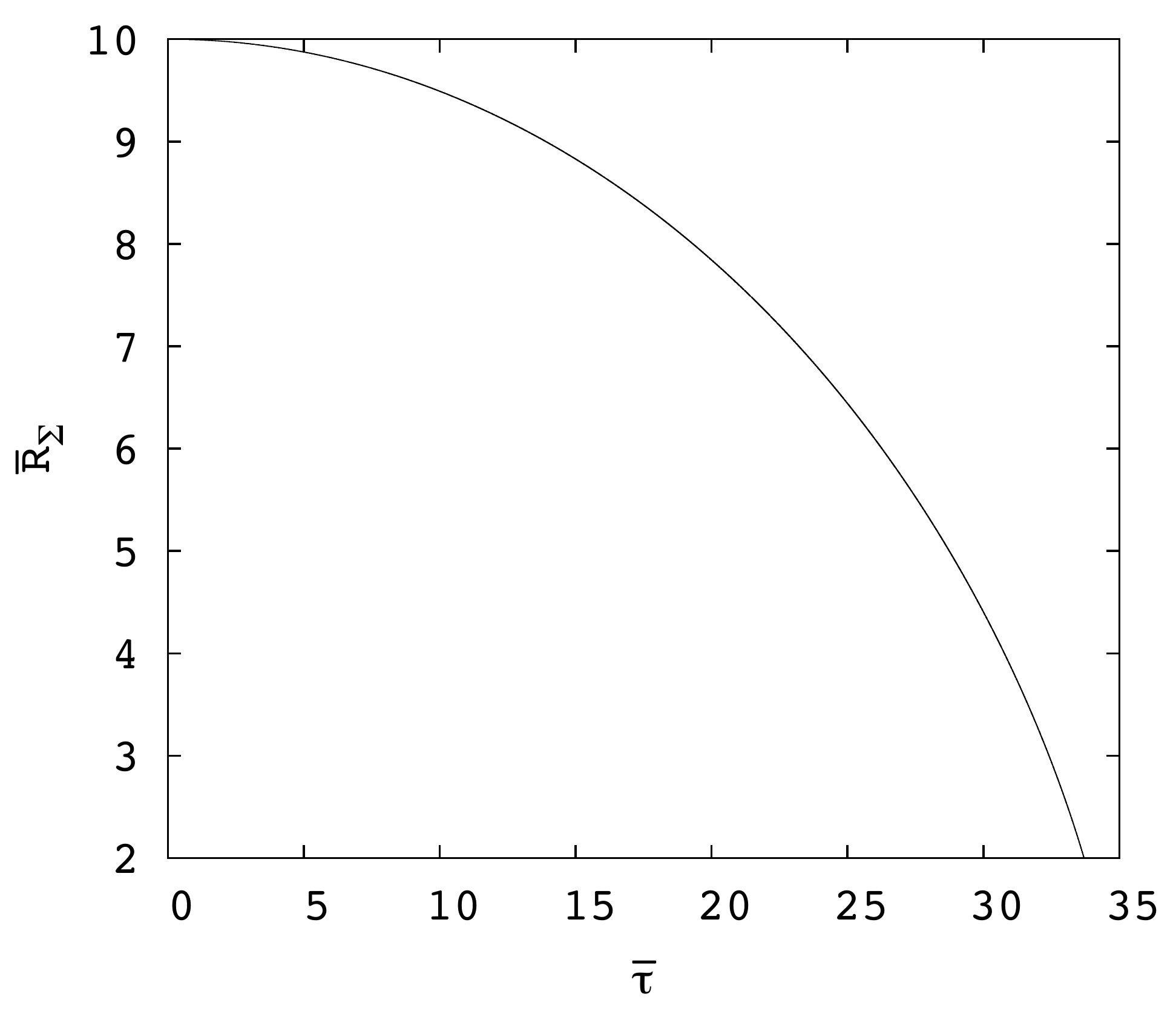}}
\caption{Evolution of the radius $\bar R_\Sigma$ for model A and initial conditions $\bar R_\Sigma(0)=10;\,\,\bar m_\Sigma(0)=1;\,\, U_\Sigma(0)=0;\,\, \displaystyle{\frac{dU_\Sigma}{d\bar\tau}}(0)=-10^{-2}$.}
\end{center}
\label{fig:figure1}
\end{figure}
\begin{figure}[htbp!]
\begin{center}
\scalebox{0.4}{\includegraphics[angle=0]{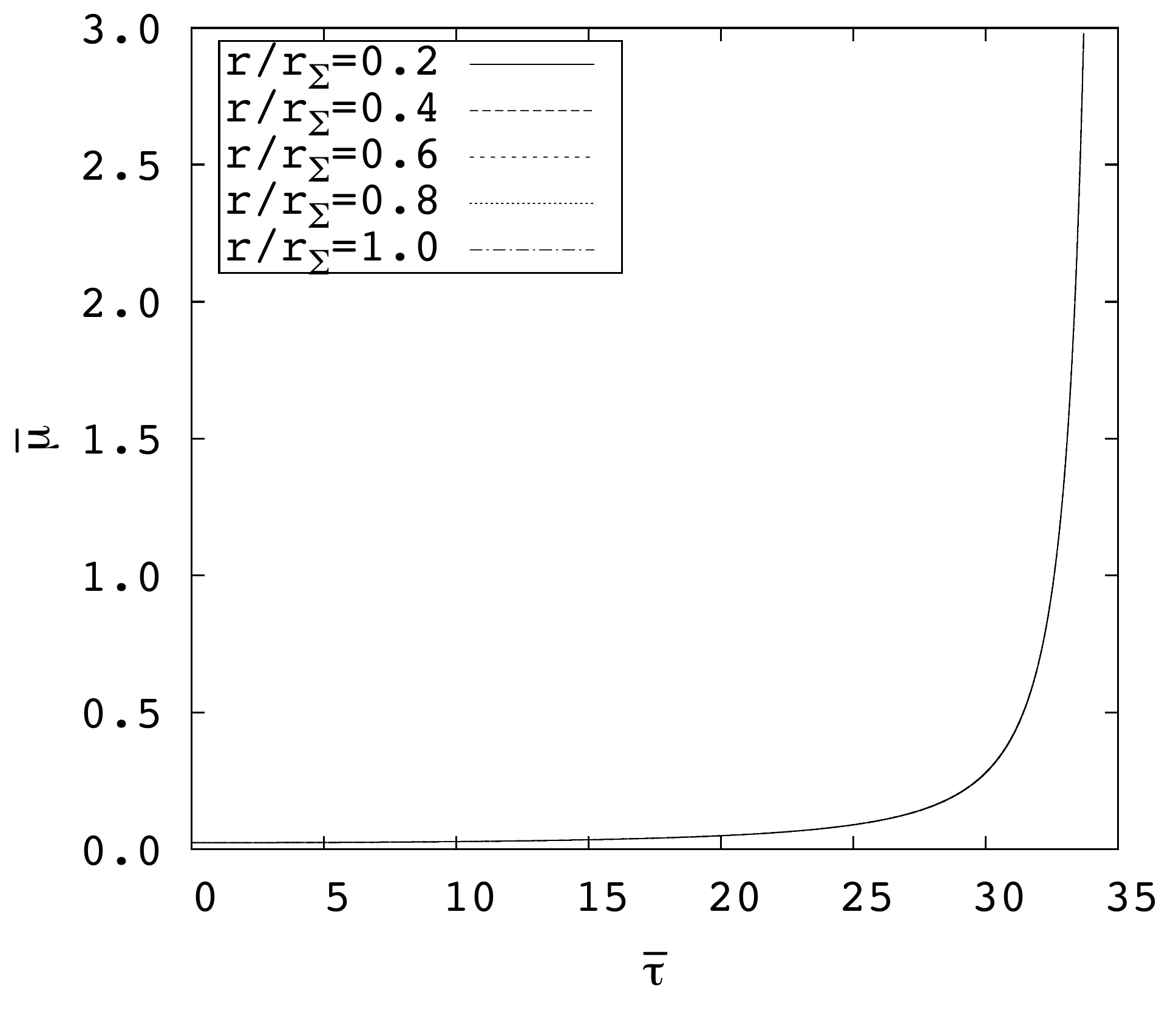}}
\caption{Evolution of the energy density $\bar\mu=m_\Sigma^2(0)\mu $ (multiplied by $10^2$) for model A and initial conditions $\bar R_\Sigma(0)=10;\,\,\bar m_\Sigma(0)=1;\,\, U_\Sigma(0)=0;\,\, \displaystyle{\frac{dU_\Sigma}{d\bar\tau}}(0)=-10^{-2}$.}
\end{center}
\label{fig:figure2}
\end{figure}
\begin{figure}[htbp!]
\begin{center}
\scalebox{0.4}{\includegraphics[angle=0]{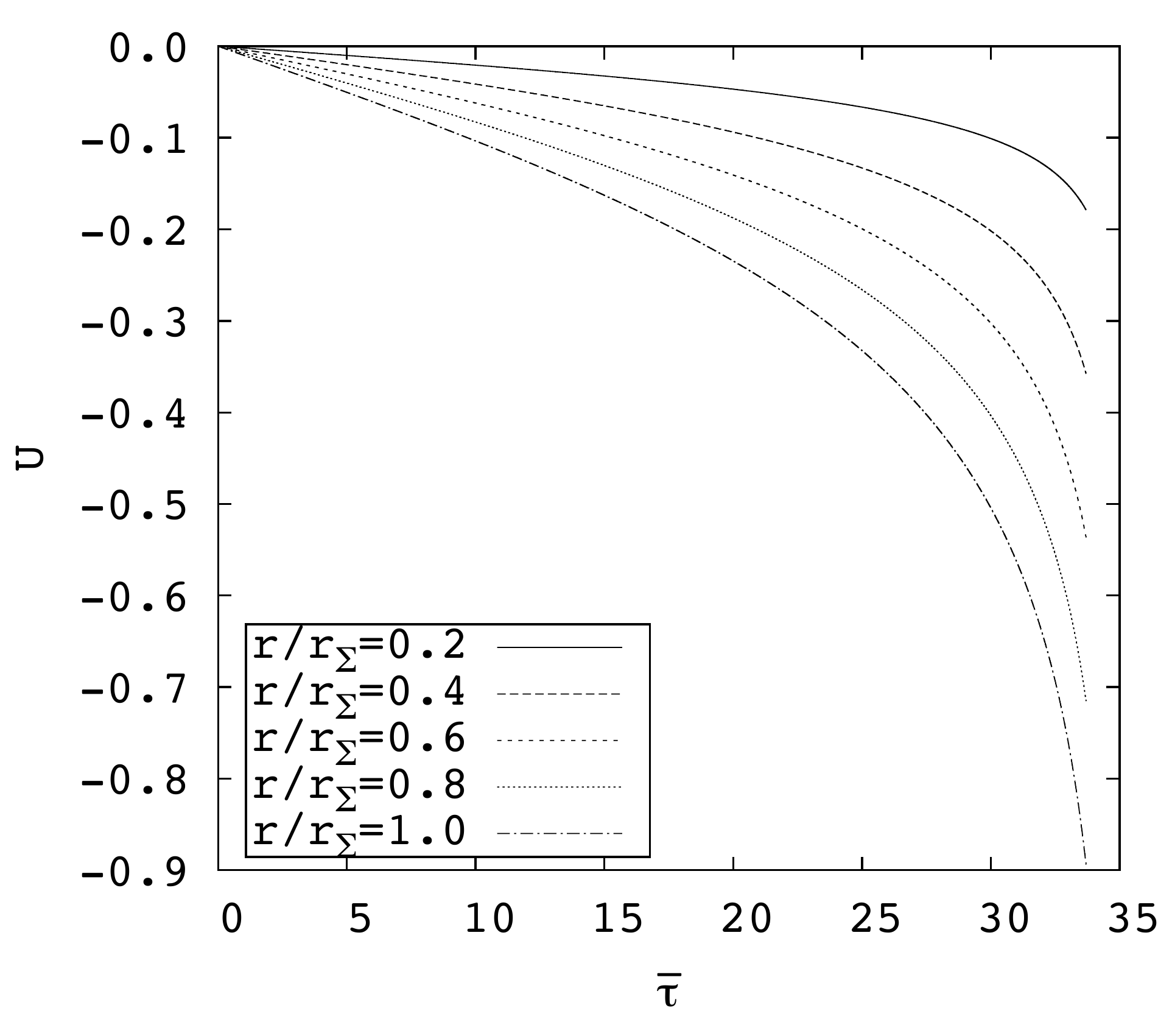}}
\caption{Evolution of the velocity $U$ for model A and initial conditions $\bar R_\Sigma(0)=10;\,\,\bar m_\Sigma(0)=1;\,\, U_\Sigma(0)=0;\,\, \displaystyle{\frac{dU_\Sigma}{d\bar\tau}}(0)=-10^{-2}$.}
\end{center}
\label{fig:figure3}
\end{figure}

\begin{figure}[htbp!]
\begin{center}
\scalebox{0.4}{\includegraphics[angle=0]{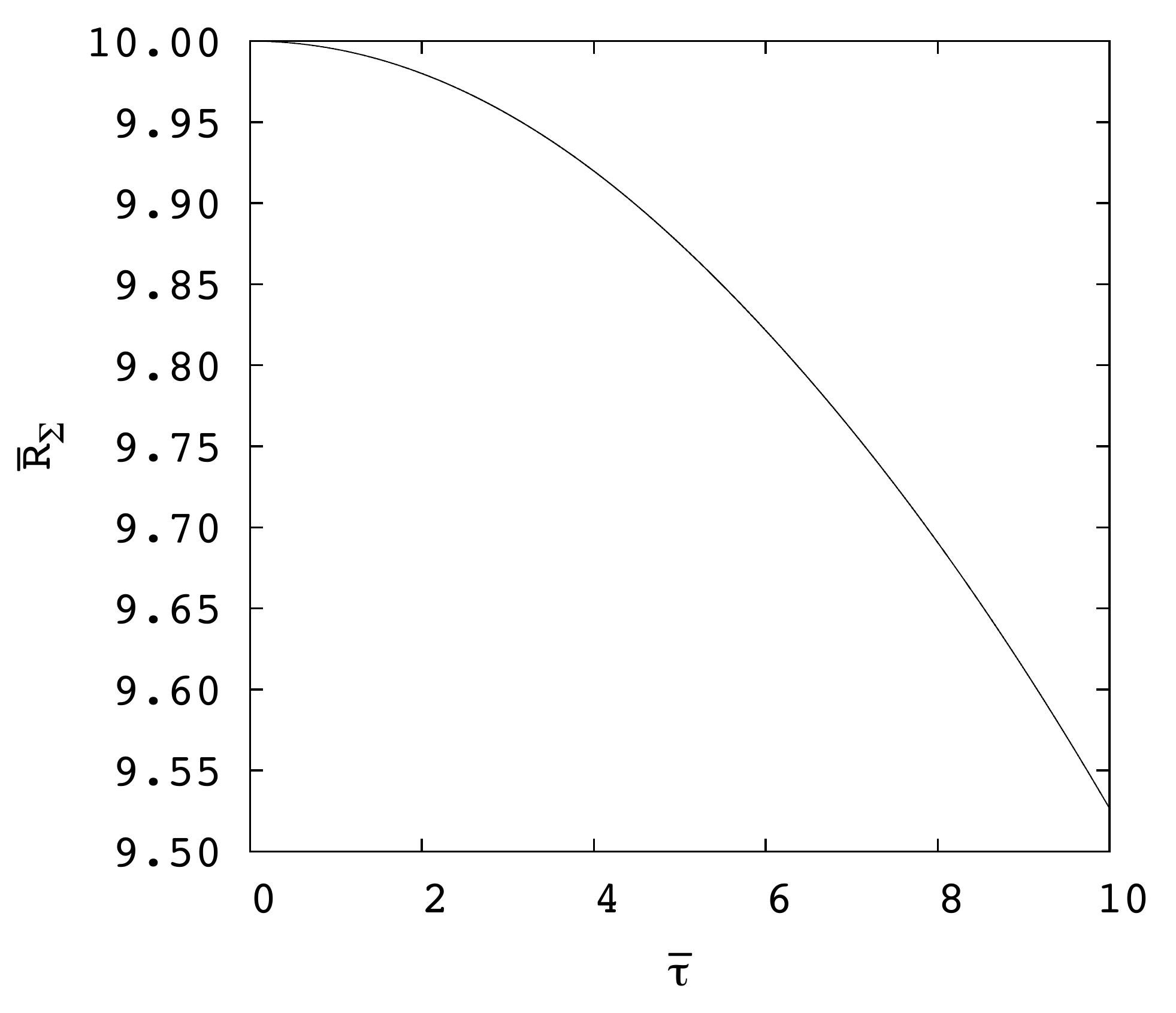}}
\caption{Evolution of the radius $\bar R_\Sigma$ for model C and initial conditions $\bar R_\Sigma(0)=10;\,\,\bar m_\Sigma(0)=1;\,\, U_\Sigma(0)=0;\,\, \displaystyle{\frac{dU_\Sigma}{d\bar\tau}}(0)=-10^{-2}$.}
\end{center}
\label{fig:figure4}
\end{figure}

\begin{figure}[htbp!]
\begin{center}
\scalebox{0.4}{\includegraphics[angle=0]{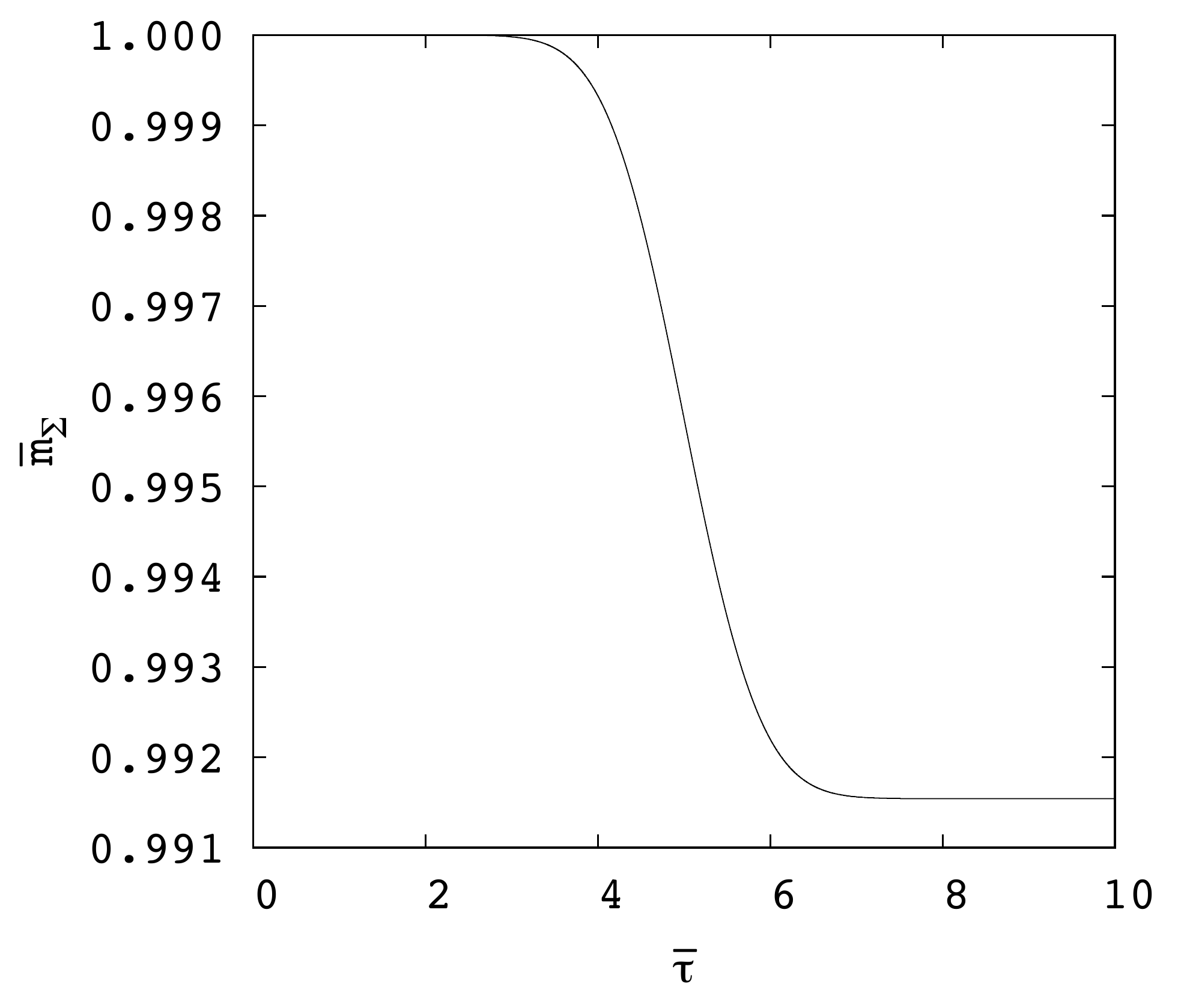}}
\caption{Evolution of the total mass $\bar m_\Sigma$ for model C and initial conditions $\bar R_\Sigma(0)=10;\,\,\bar m_\Sigma(0)=1;\,\, U_\Sigma(0)=0;\,\, \displaystyle{\frac{dU_\Sigma}{d\bar\tau}}(0)=-10^{-2}$.}
\end{center}
\label{fig:figure5}
\end{figure}

\begin{figure}[htbp!]
\begin{center}
\scalebox{0.4}{\includegraphics[angle=0]{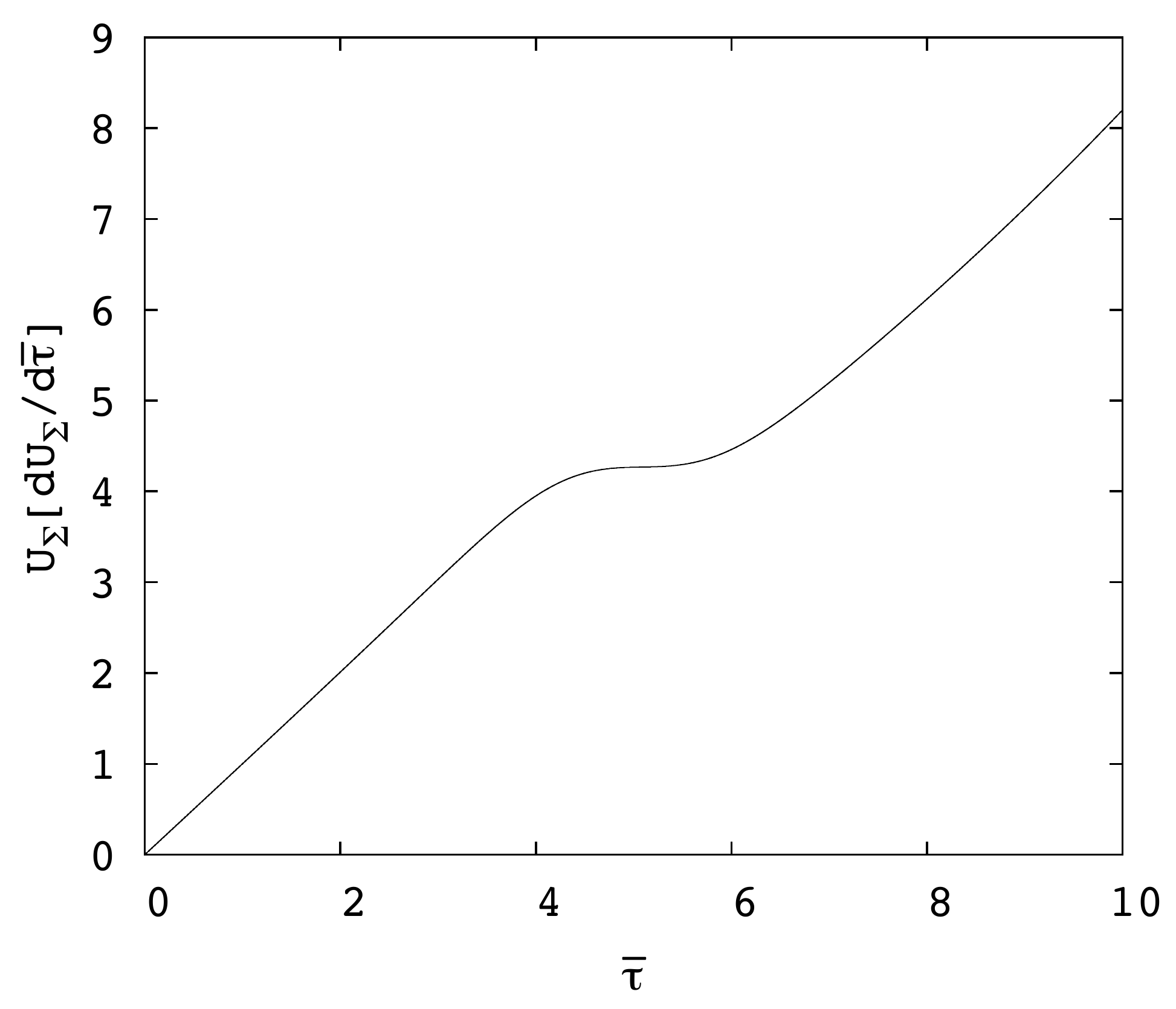}}
\caption{Evolution of $U_\Sigma [dU_\Sigma/d\bar\tau]$ (multiplied by $10^4$) for model C and initial conditions $\bar R_\Sigma(0)=10;\,\,\bar m_\Sigma(0)=1;\,\, U_\Sigma(0)=0;\,\, \displaystyle{\frac{dU_\Sigma}{d\bar\tau}}(0)=-10^{-2}$.}
\end{center}
\label{fig:figure6}
\end{figure}

\section{Protocol}
On the basis of all comments above, we shall now present the following algorithm:
\begin{enumerate}
\item Take an interior  (```seed'') solution to Einstein equations, representing a fluid distribution of matter in equilibrium, with a given 

$$\mu=\mu(r);\,\qquad\, P_{r}= P_{r}(r).$$

\item Assume that the $r$ dependence of effective variables  is the same as that of $P_{r}$ and $\mu$ respectively  and $R=r\kappa(t)$.  This assures that the remaining metric functions share the same radial dependence as that of the ``seed'' solution.

\item Using equations (\ref{meff}) and (\ref{AE}), with the $r$ dependence of $ P_{eff}$ and $\mu_{eff}$, one gets $m_{eff}$ and $A$ up to some functions of $t$.

\item For these functions of $t$ one has three ordinary differential equations (hereafter referred to as Surface equations), namely:
\begin{enumerate}
\item Equation (\ref{19}) evaluated  on $r=r_{\Sigma}$.

\item Equation (\ref{22Dt}) evaluated on $r=r_{\Sigma}$.

\item Equation  (\ref{3m}) evaluated on $r=r_{\Sigma}$.
\end{enumerate}
\item  Depending on the kind of matter under consideration, the system of surface equations described above may be closed with the additional information provided by the transport equation
and/or the equation of state for the anisotropic pressure and/or eventual additional information about some of the physical variables evaluated on the boundary surface (e.g. the
luminosity).

\item Once the system of Surface equation is closed, it may be integrated for any particular initial data.

\item From the result   of integration we obtain $R$ and then using  (\ref{20x}), (\ref{meff})  and (\ref{AE}),  $m_{eff}$, $B$ and $A$,   are completely determined.

\item With the input from the point 7 above, and using field equations, together with the equations of state and/or transport equation, all physical variables may be found for any piece of
matter distribution.
If the system is ``very far'' from equilibrium then it  could  be necessary to go  through the process once again by replacing the seed solution by the solution obtained  from the point 7 above. This could be done as many  times as it is required to obtain a satisfactory description of the system under consideration.
\end{enumerate}

Let us now elaborate on the surface equations,  which are the corner stones of the proposed algorithm.

As mentioned above, the first surface equation is  equation (\ref{19}) evaluated  on $r=r_{\Sigma}$, i.e.
\begin{equation}
\dot R_{\Sigma}=U_{\Sigma}A_{\Sigma},
\label{sur1}
\end{equation}
or 
using (\ref{junction1f})
\begin{equation}
\frac{d R_\Sigma}{d\tau}=U_\Sigma.
\label{sur2}
\end{equation}

It would be convenient to introduce the dimensionless variables
\begin{equation} 
\bar t=\frac{t}{m_{\Sigma}(0)},
\label{sur3}
\end{equation}

\begin{equation} 
\bar \tau=\frac{\tau}{m_{\Sigma}(0)},
\label{sur4}
\end{equation}
\begin{equation} 
\bar m_\Sigma=\frac{m_\Sigma}{m_{\Sigma}(0)},
\label{sur4bis}
\end{equation}
and 
\begin{equation} 
\bar R_\Sigma=\frac{R_\Sigma}{m_{\Sigma}(0)},
\label{sur5}
\end{equation}
where  $m_{\Sigma}(0)$ denotes the total initial mass. 

Then, the first surface equation reads

\begin{equation}
\frac{d\bar R_\Sigma}{d\bar t}=U_\Sigma A_\Sigma, \label{sur6}
\end{equation}
or
\begin{equation}
\frac{d\bar R_\Sigma}{d\bar \tau}=U_\Sigma.  \label{sur7}
\end{equation}

Next, evaluating  (\ref{22Dt}) on $\Sigma$ and using  (\ref{j3}), we get 
\begin{equation}
D_Tm\stackrel{\Sigma}{=} -4\pi\left[(q+\epsilon)(U+E)\right] R^2,\label{sur8}
\end{equation}
or using  (\ref{junction1f2}) and (\ref{20lum})
\begin{equation}
D_Tm\stackrel{\Sigma}{=} -(U+E) L.\label{sur9}
\end{equation}
In terms of  dimensionless variables  this last equation reads
\begin{equation}
\frac{d\bar m_\Sigma}{d\bar \tau}= -(U+E)_\Sigma L_\Sigma,
\label{sur10}
\end{equation}
or
\begin{equation}
\frac{d\bar m_\Sigma}{d\bar t}= -A_\Sigma(U+E)_\Sigma L_\Sigma.
\label{sur11}
\end{equation}
\\
Equation (\ref{sur10}) (or (\ref{sur11})) is the second surface equation. Instead of working with  $L_\Sigma$ (the luminosity of the object as measured on its surface), we may replace it by the luminosity measured by an observer at infinity $L_\infty$, using (\ref{ju}).

The third surface equation may be obtained from the ``Euler equation'' (\ref{3m})  evaluated at the boundary surface. The general form of this equation is quite long, and we shall only write it explicitly for the simplified models considered below.
\begin{figure}[htbp!]
\begin{center}
\scalebox{0.4}{\includegraphics[angle=0]{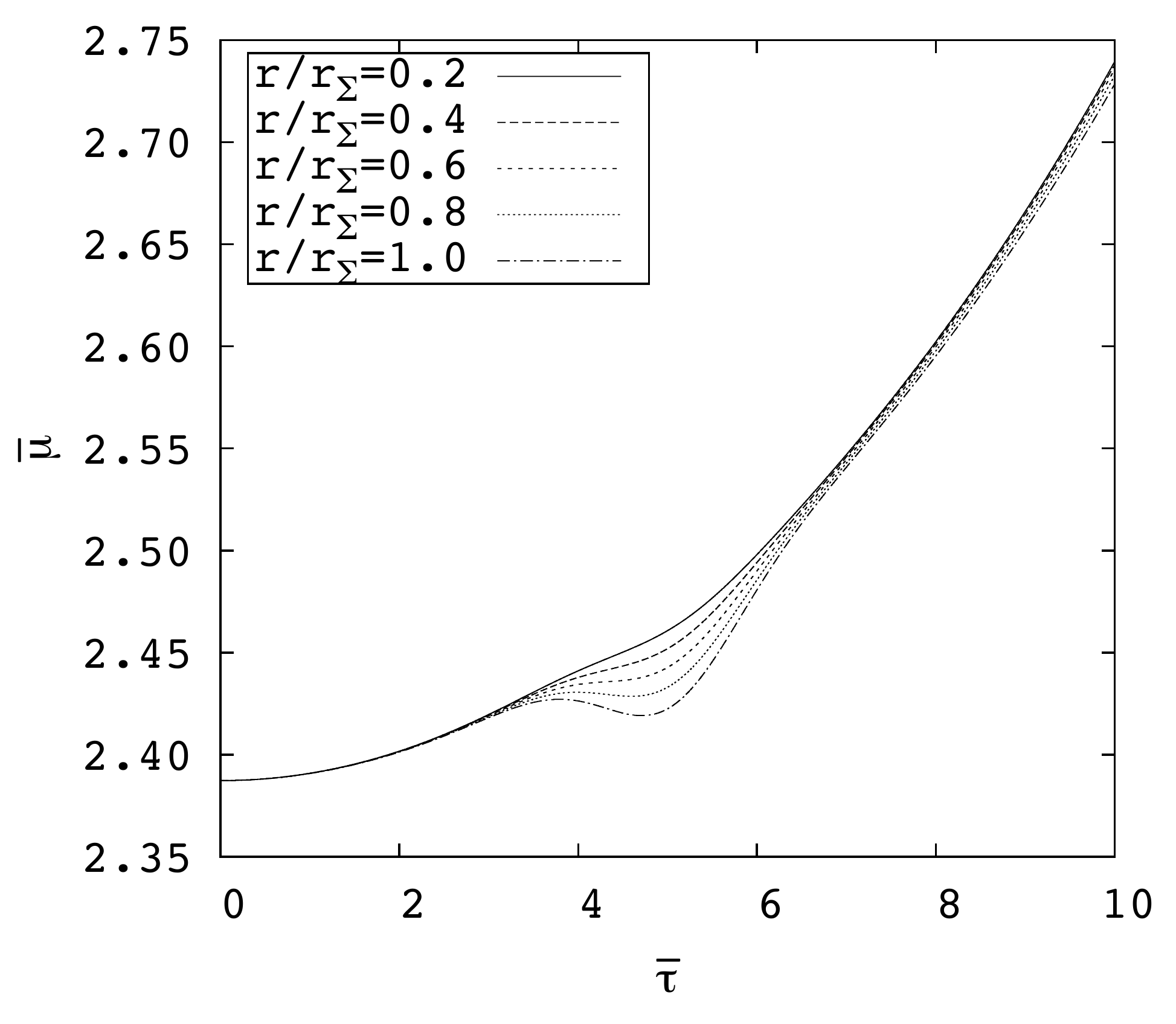}}
\caption{Evolution of the energy density $\bar\mu=m_\Sigma^2(0)\mu$ (multiplied by $10^4$) for model C and initial conditions $\bar R_\Sigma(0)=10;\,\,\bar m_\Sigma(0)=1;\,\, U_\Sigma(0)=0;\,\, \displaystyle{\frac{dU_\Sigma}{d\bar\tau}}(0)=-10^{-2}$.}
\end{center}
\label{fig:figure7}
\end{figure}
\begin{figure}[htbp!]
\begin{center}
\scalebox{0.4}{\includegraphics[angle=0]{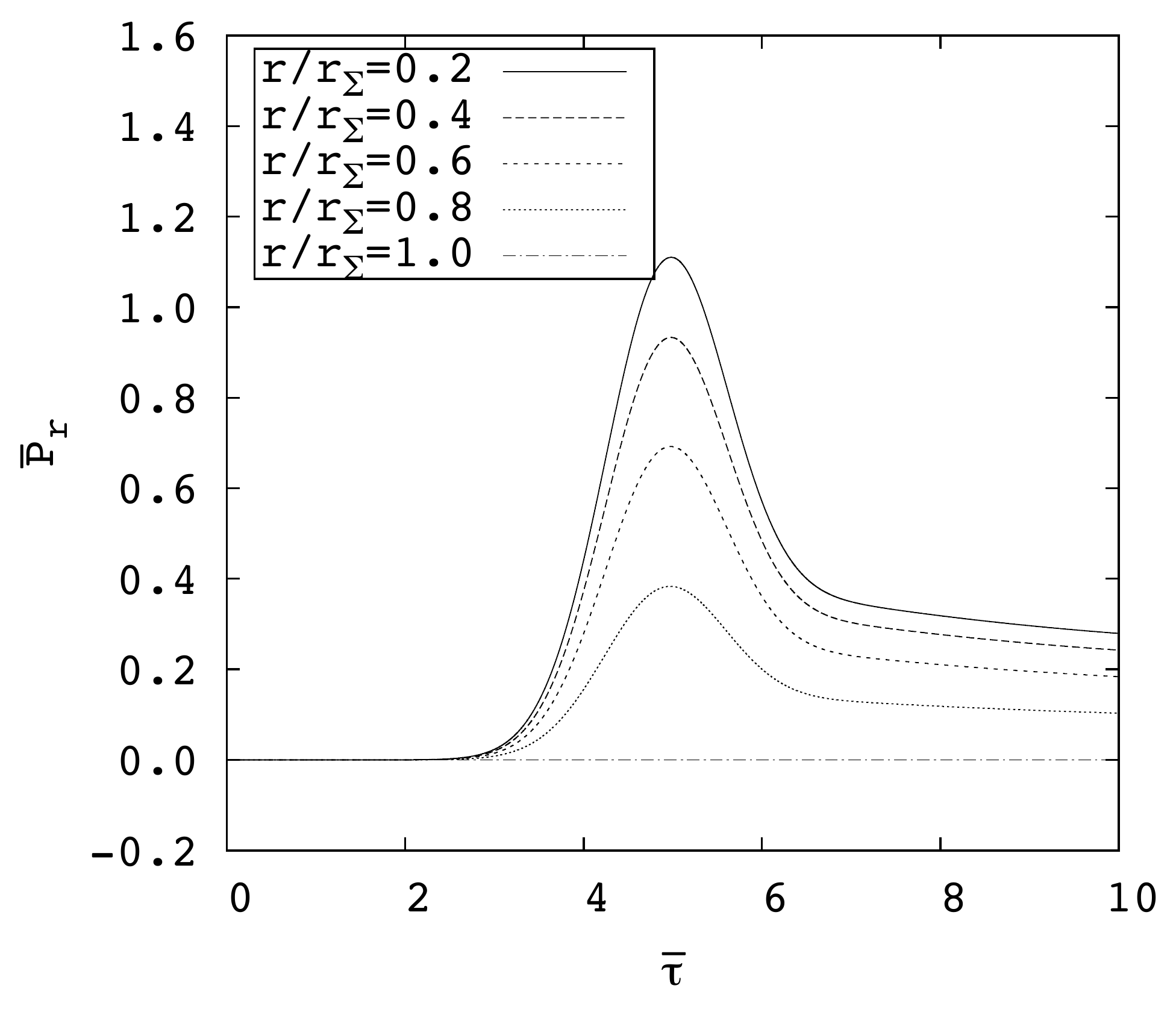}}
\caption{Evolution of the radial pressure $\bar P_r=m_\Sigma^2(0)P_r$ (multiplied by $10^5$) for model C and initial conditions $\bar R_\Sigma(0)=10;\,\,\bar m_\Sigma(0)=1;\,\, U_\Sigma(0)=0;\,\, \displaystyle{\frac{dU_\Sigma}{d\bar\tau}}(0)=-10^{-2}$.}
\end{center}
\label{fig:figure8}
\end{figure}
\begin{figure}[htbp!]
\begin{center}
\scalebox{0.4}{\includegraphics[angle=0]{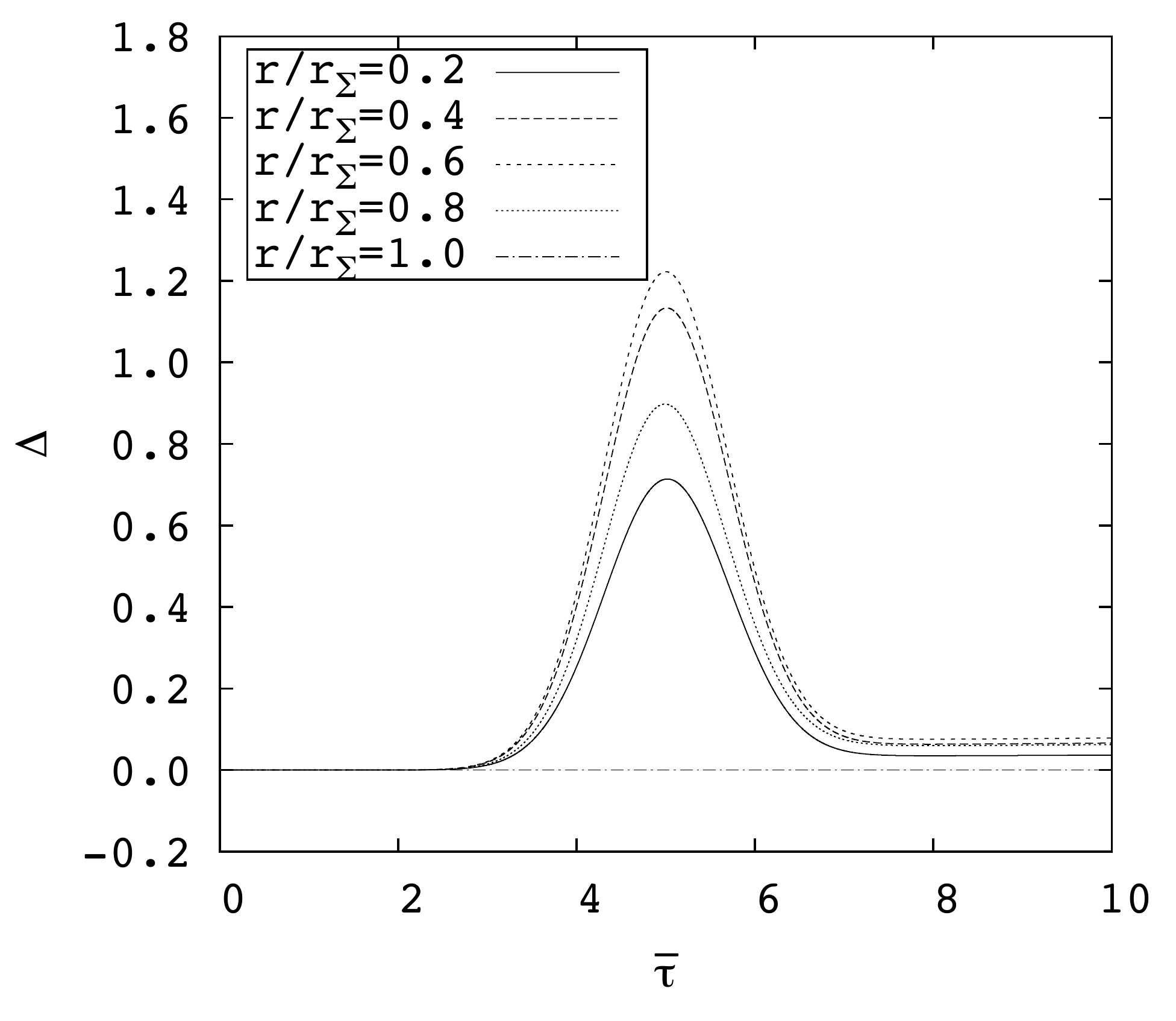}}
\caption{Evolution of the grade of anisotropy $\Delta=\bar P_\perp-\bar P_r=m_\Sigma^2(0)(P_\perp-P_r)$ (multiplied by $10^6$) for model C and initial conditions $\bar R_\Sigma(0)=10;\,\,\bar m_\Sigma(0)=1;\,\, U_\Sigma(0)=0;\,\, \displaystyle{\frac{dU_\Sigma}{d\bar\tau}}(0)=-10^{-2}$.}
\end{center}
\label{fig:figure9}
\end{figure}
\begin{figure}[htbp!]
\begin{center}
\scalebox{0.4}{\includegraphics[angle=0]{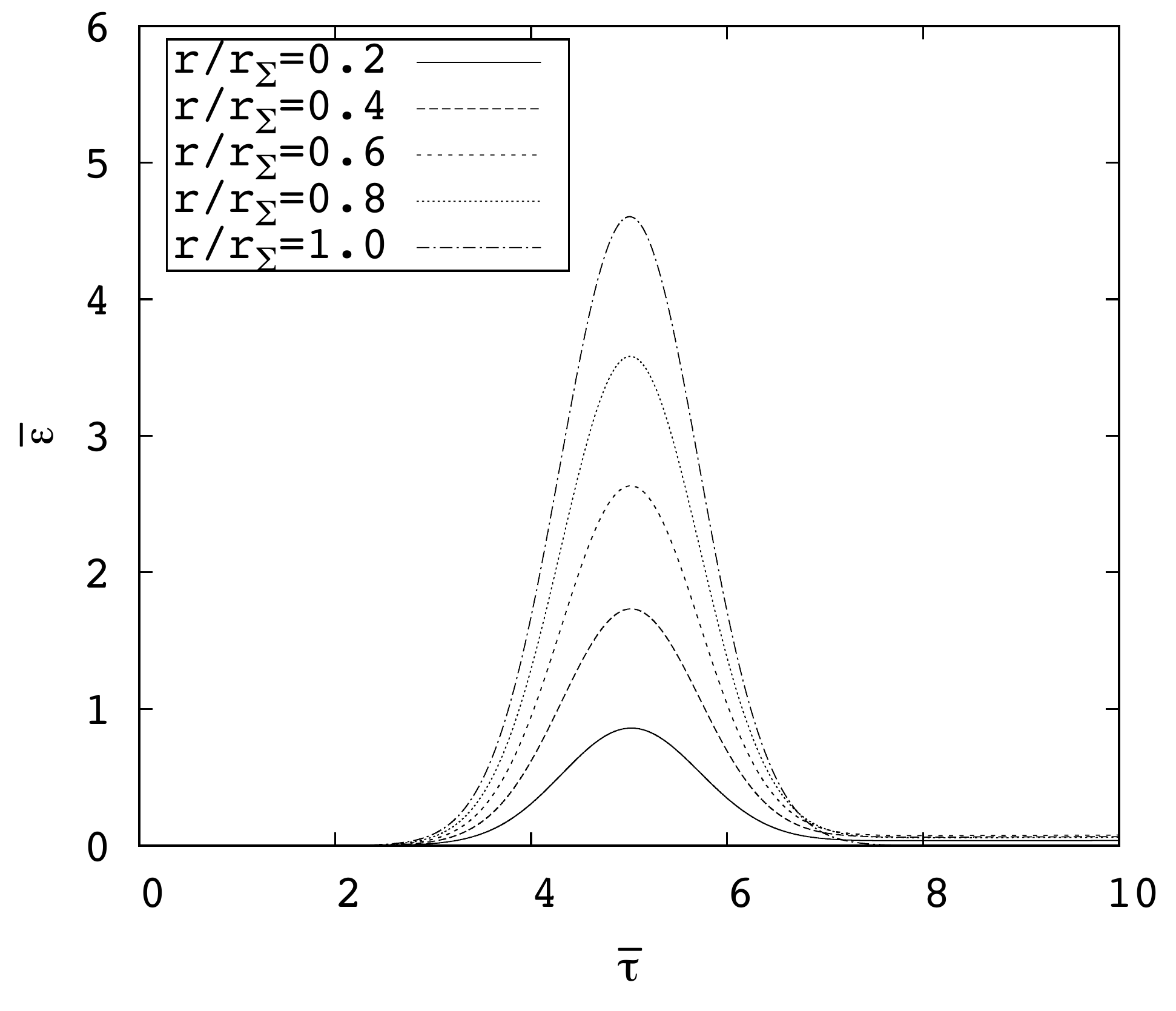}}
\caption{Evolution of the energy flux $\bar\epsilon=m_\Sigma^2(0)\epsilon$ (multiplied by $10^6$) for model C and initial conditions $\bar R_\Sigma(0)=10;\,\,\bar m_\Sigma(0)=1;\,\, U_\Sigma(0)=0;\,\, \displaystyle{\frac{dU_\Sigma}{d\bar\tau}}(0)=-10^{-2}$.}
\end{center}
\label{fig:figure10}
\end{figure}
\begin{figure}[htbp!]
\begin{center}
\scalebox{0.4}{\includegraphics[angle=0]{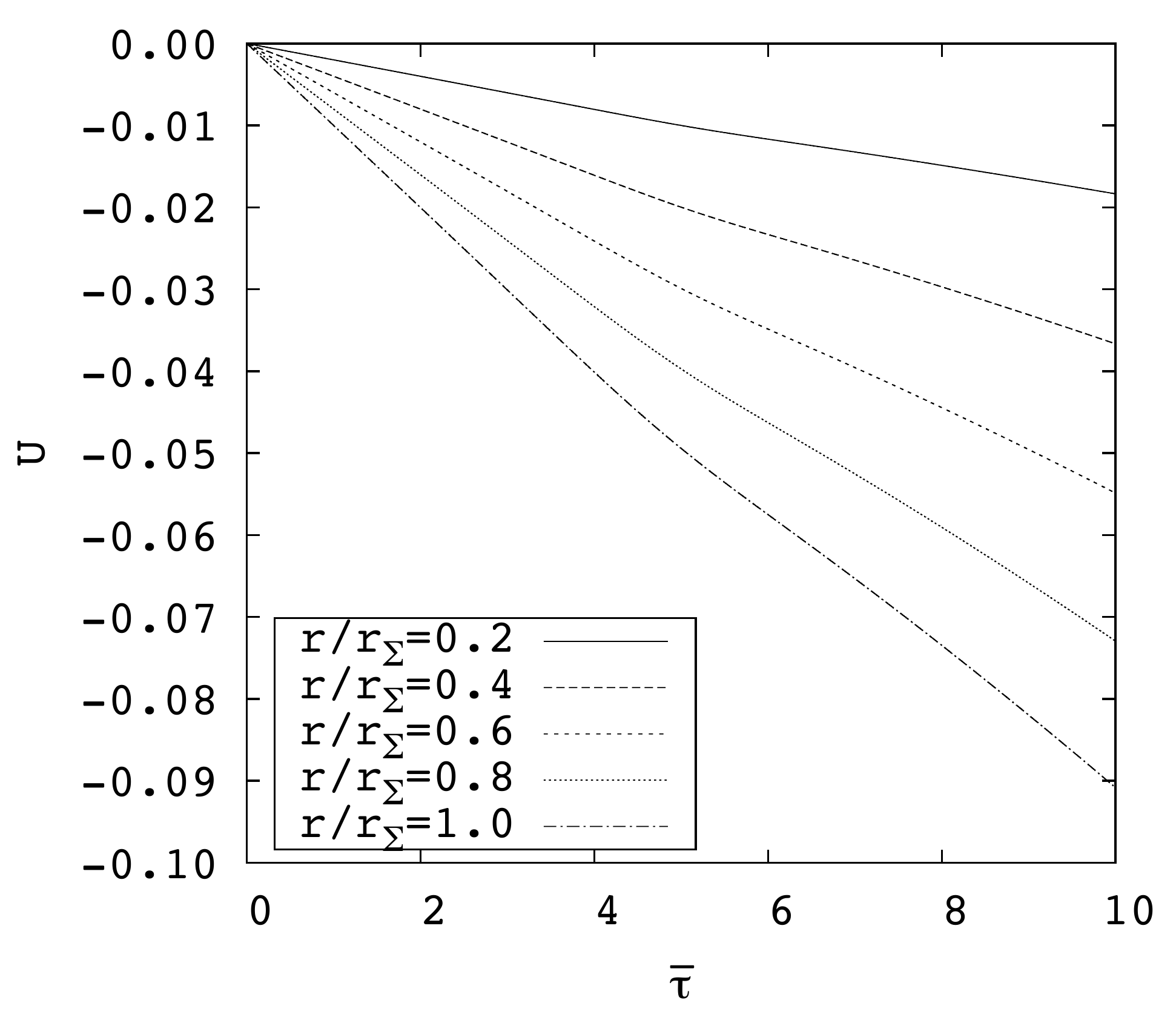}}
\caption{Evolution of the velocity $U$ for model C and initial conditions $\bar R_\Sigma(0)=10;\,\,\bar m_\Sigma(0)=1;\,\, U_\Sigma(0)=0;\,\, \displaystyle{\frac{dU_\Sigma}{d\bar\tau}}(0)=-10^{-2}$.}
\end{center}
\label{fig:figure11}
\end{figure}
\section{Modeling: Some examples}
We shall now illustrate the method outlined above with some examples inspired in the well known Schwarzschild interior solution. It should be clear that our goal here is not to solve any specific astrophysical problem, but just to exhibit the potential of the method. Also it should be mentioned that models inspired in the Schwarzschild interior solution (within the context of the PQSR) were presented in \cite{HJR}, \cite{hjr2} and \cite{hbds02}, using the version of the method in non--comoving coordinates. However, the initial data as well as the type of anisotropy used here differ from the one assumed for those previous models,  and accordingly the evolution pattern is quite different. This latter  point strees further the fact that in spite of the similarity of the basic assumption underlying the very definition of the PQSR, the framework of both versions (in comoving and non--comoving frames) are are very  different.

For the sake of simplicity we shall consider nonviscous fluid, dissipating  only in the
streaming out limit (i.e. $q=\eta=0$).

Thus, the  effective density for the Schwarzshild type model is assumed  to depend only on the time--like coordinate, i.e.
\begin{equation}\mu_{eff}=f(t) \label{n1r},\end{equation}
whereas the expression  for the effective pressure is
\begin{equation}
\frac{P_{eff}+\frac{1}{3}\mu_{eff}}{\mu_{eff} + P_{eff}}=\xi\lambda(t),\label{sin1}
\end{equation}
with
\begin{equation}
\xi=\left(1-\frac{8\pi}{3} \mu_{eff} R^2\right)^{1/2}.
\end{equation}
We may also write (\ref{sin1}) as 
\begin{equation}
P_{eff}=\mu_{eff}\frac{(3\xi\phi-\chi)}{3(\chi-\xi\phi)} \label{nr2},
\end{equation}
where  $\lambda=\phi/\chi$ is an arbitrary function  of time such that the pressure
satisfy the boundary condition, which in this case is $P_r\stackrel{\Sigma}{=}0$.

It is easy to check that
\begin{equation}
\phi={\bar R_\Sigma L_\Sigma+\bar m_\Sigma+\displaystyle{\frac{dU_\Sigma}{d\bar\tau}} \bar R_\Sigma^2},
\end{equation}
\begin{equation}
\chi={E_\Sigma\left\{\bar R_\Sigma L_\Sigma+3\bar m_\Sigma-\bar R_\Sigma U_\Sigma^2
+\displaystyle{\frac{dU_\Sigma}{d\bar\tau}} \bar R_\Sigma^2\right\}}.
\end{equation}

Also, using (\ref{meff}) and (\ref{AE}) we obtain
\begin{equation}
m_{eff}=\frac{4\pi}{3}R^3 f,
\end{equation}
\begin{equation}
A=A_\Sigma\frac{\Psi}{\Psi_\Sigma},
\label{sol}
\end{equation}
where
\begin{equation}
\Psi=\chi-\xi\phi \label{puya},
\end{equation}
and
\begin{equation}
\Psi_\Sigma=E_\Sigma \bar R_\Sigma\left(\frac{2\bar m_\Sigma}{\bar R_\Sigma}-U_\Sigma^2\right).
\label{puyas}
\end{equation}
Then (\ref{3m}) is written as

\begin{eqnarray}
&&(\tilde\mu+\tilde P_r)D_T U + (\tilde\mu+\tilde P_r)\left(4\pi R \tilde P_r + \frac{m}{R^2}\right)\nonumber \\
&& + E^2 \left(D_R \tilde P_r-\frac{2}{R}P_\perp\right)\nonumber\\
&& + E\left[D_T\epsilon + 2\epsilon\left(\frac{2U}{R}+\sigma\right)\right]=0.
\label{n3m}
\end{eqnarray}

Next, we shall evaluate (\ref{n3m}) at the boundary  surface, for doing that we shall need an equation of state for the stresses (at the boundary). For the sake of simplicity we consider $[P_\perp]_\Sigma=0$,  then after lenghty manipulations using MAPLE, we are lead  straightforwardly to the third equation
at the surface
\begin{equation}
U_\Sigma \frac{d^2U_\Sigma}{d\bar\tau^2}= \mathcal{A}\left(\frac{dU_\Sigma}{d\bar\tau}\right)^2
+\mathcal{B}\frac{dU_\Sigma}{d\bar\tau}+\mathcal{C},\label{TE}
\end{equation}
where
\begin{equation}
\mathcal{A}=\frac{1}{E_\Sigma^2}\left(\frac{2\bar m_\Sigma}{\bar R_\Sigma}+2U_\Sigma^2-1\right),
\end{equation}

\begin{eqnarray}
\mathcal{B}&=&\frac{1}{\bar R_\Sigma^4 E_\Sigma^2}\left(-3U_\Sigma^4 \bar R_\Sigma^3-3U_\Sigma^2\bar R_\Sigma^3+11U_\Sigma^2\bar m_\Sigma \bar R_\Sigma^2\right.\nonumber\\
&+&\left. 2\bar m_\Sigma^2\bar R_\Sigma-\bar m_\Sigma \bar R_\Sigma^2 +[D_Tm]_\Sigma U_\Sigma \bar R_\Sigma^3\right),
\end{eqnarray}

\begin{eqnarray}
\mathcal{C}&=&\frac{1}{\bar R_\Sigma^4 E_\Sigma^2}[L_\Sigma(2\bar R_\Sigma^2+2U_\Sigma^4 \bar R_\Sigma^2+
8\bar m_\Sigma^2 -8U_\Sigma^2 \bar R_\Sigma \bar m_\Sigma \nonumber\\
&&+ 4 U_\Sigma^2 \bar R_\Sigma^2 -8\bar R_\Sigma \bar m_\Sigma) + [D_Tm]_\Sigma\left(8 \bar m_\Sigma \bar R_\Sigma U_\Sigma \right.
\nonumber\\
&&\left.-\bar R_\Sigma^2 U_\Sigma^3 - \bar R_\Sigma^2 U_\Sigma -3 [D_Tm]_\Sigma \bar R_\Sigma^2\right)\nonumber \\ 
&&-U_\Sigma^4\bar R_\Sigma \bar m_\Sigma+5U_\Sigma^2\bar m_\Sigma^2-\bar R_\Sigma U_\Sigma^2 \bar m_\Sigma].
\end{eqnarray}

The system of surface equations (\ref{sur7}), (\ref{sur10}) and (\ref{TE}) has been  integrated  for the following \ initial conditions

$$\bar R_\Sigma(0)=10;\,\,\bar m_\Sigma(0)=1;\,\, U_\Sigma(0)=0;\,\, \frac{dU_\Sigma}{d\bar\tau}(0)=-10^{-2}$$

We shall consider  both, the adiabatic and the dissipative case (in the free streaming approximation), in the latter case  we shall use a Gaussian as a luminosity profile
$$L_\Sigma=L_0 e^{-(\bar\tau-\bar\tau_0)^2/\xi}.$$
We integrate numerically the system of
equations at the surface using a standard fourth order Runge--Kutta. We built up
a nonadiabatic model preparing as test beds two previous models do not deprived
of physical interest.

 For simplicity we have cut  here the iterative process after  the first step of the chain (i.e. we have assumed the seed solution as given by (\ref{n1r}) and (\ref{nr2})). We invite any interested reader to proceed to the next step of the iterative method by replacing the effective variables by the energy density and pressure obtained here in the first step and go through the protocol once again.

\subsection{Model A: Fully adiabatic}
In this model we use $D_T B=BD_R U$ everywhere which is equivalent to enforce $\epsilon=0$. Thus, setting $L_0=0$ we obtain results displayed in figures 1--3. The isotropic fluid sphere behaves as dust, that is $P_r=P_\perp=0$ everywhere, and collapse proceeds catastrophically 
as expected. The monitoring of the evolution is stopped when trespassing the horizon.

\subsection{Model B: Without luminosity}
We can set only $L_0=0$ to see how the distribution evolves. With no surprise
the results are the same as in model A. This is a consequence of the well--posed
initial--boundary problem.

\subsection{Model C: Nonadiabatic}
When some fraction of the total mass ($\approx 1\%$) is carried away by the Gaussian pulse, picked at $\bar \tau_0=5$, with $\xi=1$, the dust ball initially goes to collapse, becoming anisotropic in the process. These results are shown in figures 4--11.

\section{Concluding remarks}
A seminumerical method  to describe gravitational collapse in  comoving coordinates has been proposed, in analogy  with the already existing algorithm in non--comoving coordinates. 

For doing that, we have revisited and redefined the basic concepts of the post--quasistatic approximation (PQSA) in order to adapt them to comoving coordinates. The essential features of the seminumeric method keep going. But in comoving coordinates we were enforced to reasonable transfer the PQSA to a geometrical variable ($R$). Up to now the effective variables let us to make heuristically the job. In the present version an additional  geometrical point of view led straightforwardly to the new effective variables. We endeavor supposing there is life in between quasistatic and post--quasistatic regimes. So we did. Here we reported one version of the PQSA in comoving coordinates.

We have integrated the surface equations for some simple models inspired in the interior Schwarzschild solution. Our intention presenting such models was not to describe any physically relevant astrophysical scenario, but just to illustrate the method.
Due to the simplifications imposed on the models, the field equations  are overdetermined producing specific constraints.  Accordingly, a fine tuning specification of initial conditions is necessary. Even more, in our models we never recover the quasistatic o the static regimes. Our models are intrinsically unstable and physically acceptable. We have to mention that we observe certain tendency to stabilize the system if the configuration initially
was less relativistic (less compact). 

Emission of energy seems to play a crucial role 
in the process of gravitational collapse. Eventually dissipation and anisotropy (unequal stresses) may change the evolution fate,
avoiding the complete collapse to a black hole in the same hydrodynamical  time scale. Some additional work is required on this last issue, specially considering a transport equation to deal with heat flow (and/or viscosity) dissipation in the context of extended thermodynamics within the PQSA (see \cite{hjr9}, \cite{HJ} for a treatment of this problem in non-comoving coordinates).

 \end{document}